\documentclass[aip,jap,amsmath,amssymb,reprint,superscriptaddress]{revtex4-1}
\usepackage{graphicx}
\usepackage{dcolumn}
\usepackage{bm}
\usepackage{color} 
\begin{document}
\title{Tutorial: Magnetic resonance with nitrogen-vacancy centers in diamond---microwave engineering, materials science, and magnetometry}
\author{Eisuke Abe}
\email{e-abe@keio.jp}
\affiliation{Spintronics Research Center, Keio University, 3-14-1 Hiyoshi, Kohoku-ku, Yokohama 223-8522, Japan}
\author{Kento Sasaki}
\affiliation{School of Fundamental Science and Technology, Keio University, 3-14-1 Hiyoshi, Kohoku-ku, Yokohama 223-8522, Japan}
\date{\today}
\begin{abstract}
This tutorial article provides a concise and pedagogical overview on negatively-charged nitrogen-vacancy (NV) centers in diamond.
The research on the NV centers has attracted enormous attention for its application to quantum sensing, encompassing the areas of not only physics and applied physics but also chemistry, biology and life sciences.
Nonetheless, its key technical aspects can be understood from the viewpoint of magnetic resonance.
We focus on three facets of this ever-expanding research field, to which our viewpoint is especially relevant: microwave engineering, materials science, and magnetometry.
In explaining these aspects, we provide a technical basis and up-to-date technologies for the research on the NV centers.
\end{abstract}
\maketitle
\section{Introduction}

Defects in semiconductors can profoundly alter the electrical, optical and magnetic properties of the host semiconductors,
and the magnetic resonance technique has been an indispensable tool for the defect characterization.
There, the spins of the defects serve as `markers' to trace the roles of the defects.
On the other hand, recent years have witnessed intense research activities aiming to harness the defect spins themselves as device functionalities.
One prominent platform is a phosphorus donor in silicon.
The electron and nuclear spins of the donor have been proposed as quantum bits for quantum information processing,
and most advanced experiments have demonstrated readout and coherent control of both electron and nuclear spins of single donors with high fidelities.~\cite{K98,PTD+12,PTD+13}

\begin{figure*}
\begin{center}
\includegraphics{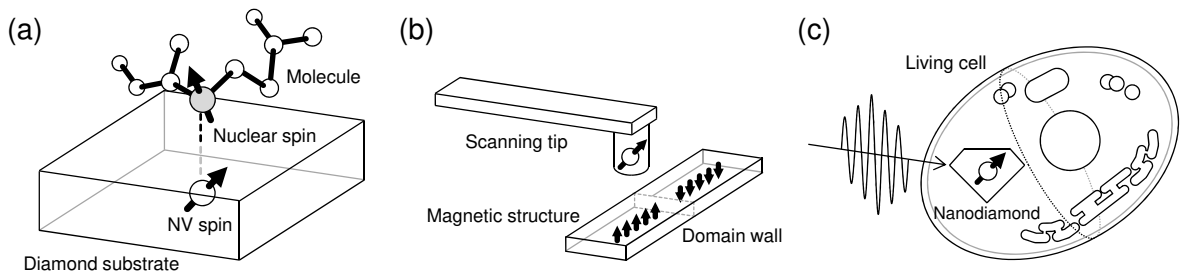}
\caption{Various modalities of NV-based magnetometers.
(a) Detection of a single nuclear spin of a single molecule using a near-surface NV center.
(b) High spatial resolution sensing of a magnetic structure using a scanning diamond probe.
(c) Nanodiamond embedded in a living cell.
\label{fig1}}
\end{center}
\end{figure*}

Another example, which we discuss in this article, is a negatively-charged nitrogen-vacancy (NV) center in diamond (hereafter referred to as NV center for short).~\cite{JW07}
Being optically addressable and coherently controllable by microwaves and preserving quantum coherence even at room temperature,
an electronic spin of the NV center can be utilized as a magnetic sensor.~\cite{MSH+08,BCK+08,SCLD14,RTH+14}
An advantage of the NV-based magnetometer lies in a variety of modalities it can take [Fig.~\ref{fig1}].
NV spins locating close to the diamond surface can detect nuclear spins of, for instance, molecules, proteins, and cells placed on top of the surface.~\cite{MKS+13,SSP+13,LSU+16}
By incorporating NV spins into a diamond-based scanning probe, an atomic-scale spatial resolution is attainable.
The scanning-type setup is being established as a new tool to probe novel magnetic structures such as domain walls and vortices in superconductors.~\cite{THK+14,TRG+16,CSY18}
Fluorescent nanodiamonds (nanometer-sized diamond particles) containing NV centers can be embedded in a living cell,
potentially allowing for magnetometry in a single cell as well as electrometry and thermometry.~\cite{DFD+11,KMY+13,NJD+13,WJPW16}
In addition, ensemble NV centers enable submicron scale, two-dimensional magnetic imaging.~\cite{LAG+13,GLP+15,SRH+17}
Clearly, the research to develop such versatile sensors is highly interdisciplinary, and various facets of science and technology must be combined together.

In this article, we focus on three aspects that are crucial in the study of the NV centers: microwave engineering, materials science, and magnetometry.
In doing so, we emphasize the pivotal role that magnetic resonance plays.
Microwave engineering is a basis for spin control.
The challenge in materials science is to sustain narrow linewidths and good coherence of the NV spins even when they are brought closer to the surface to achieve better magnetic sensitivity.
Magnetic sensing protocols largely rely on numerous pulse sequences developed in the field of magnetic resonance.

This article is organized as follows.
In Sec.~\ref{sec_basic}, basic physics of the NV center and experimental techniques for optically-detected magnetic resonance (ODMR) are outlined.
Microwave engineering is discussed in Sec.~\ref{sec_me}.
Materials science aspects of the research, N$^+$ ion implantation and nitrogen-doping during chemical vapor deposition (CVD) techniques to create NV centers, are discussed in Sec.~\ref{sec_ms}.
Sections~\ref{sec_me} and \ref{sec_ms} also serve a purpose of presenting typical experimental data in this field so that the readers from different fields can get the idea.
DC and AC magnetometries are discussed in Sec.~\ref{sec_dc} and Sec.\ref{sec_ac}, respectively.
Conclusion is given in Sec.~\ref{sec_conc}.

We note that, since this article is a tutorial and not a review, it is not meant to be exhaustive.
We give references whenever relevant, but limit them to selected ones which are most directly connected to the subjects of interest.
Rather than referring to other works, we illustrate the subjects using concrete examples from our own research activities (either previously published or unpublished).
In some cases, similar data have been obtained and published by other research groups preceding ours.

\section{Basics\label{sec_basic}}
\begin{figure}
\begin{center}
\includegraphics{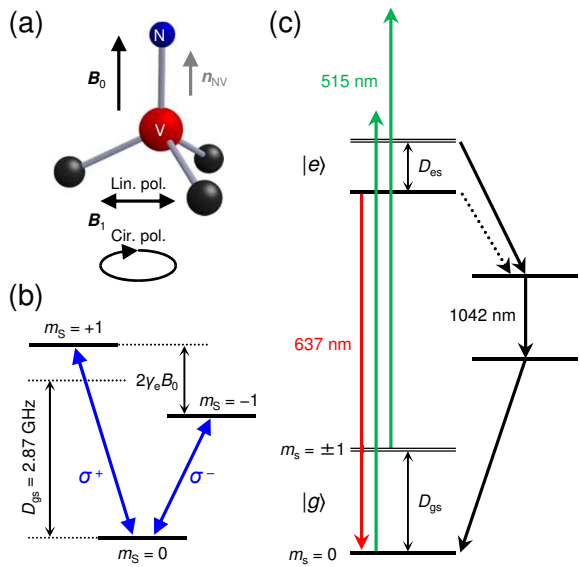}
\caption{(a) Schematic of the NV center.
(b) Spin triplet ground state of the NV spin.
(c) Energy level diagram of the NV center (not to scale).
\label{fig2}}
\end{center}
\end{figure}

The NV center in diamond is an atomic defect consisting of a substitutional nitrogen and a vacancy adjacent to it [Fig.~\ref{fig2}(a)].
The nitrogen atom and three neighboring carbon atoms provide five electrons, and an extra electron is captured to make the system negatively-charged.
The four electronic orbital states of the dangling bonds surrounding the vacancy are transformed under the C$_{3v}$ point-group symmetry.
The irreducible representations are labeled as $a'_1$, $a_1$, $e_x$, and $e_y$.
The $a'_1$ and $a_1$ states are fully occupied by two electrons in each state, and the degenerate $e_x$ and $e_y$ states accommodate one electron each.
This configuration makes the ground state $|g\rangle$ of the NV center an $S$ = 1 electronic spin triplet system.

The basic spin physics of the NV center can be captured by the Hamiltonian
\begin{equation}
\mathcal{H} = D_{\mathrm{gs}} S_z^2 + \gamma_{\mathrm{e}} B_0 S_z,
\label{eq_spin}
\end{equation}
where $D_{\mathrm{gs}}$ = 2.87~GHz is the ground state zero-field splitting,
$\gamma_{\mathrm{e}}$ = 28~GHz/T is the gyromagnetic ratio of the electronic spin.
The quantization ($z$) axis of Eq.~(\ref{eq_spin}) is defined to be along the NV axis, the vector connecting the nitrogen atom and the vacancy [Fig.~\ref{fig2}(a)],
and $S_z$ is the $z$-component of the $S$ = 1 spin operator.
$B_0$ is the static magnetic field applied along the NV axis. 
The energy levels described by Eq.~(\ref{eq_spin}) are depicted in Fig.~\ref{fig2}(b).

The NV center absorbs photons at 637~nm.
In this process, one electron in $a_1$ is promoted to either $e_x$ or $e_y$, and there is one unpaired electron left in $a_1$.
The description of this excited state $|e\rangle$ is much more complicated than that of $|g\rangle$.
But at room temperature, phonons mix some of the orbital states, making $|e\rangle$ behave as spin triplet similar to $|g\rangle$.
Equation~(\ref{eq_spin}) is also valid for $|e\rangle$ if $D_{\mathrm{gs}}$ is replaced with the excited state zero-field splitting $D_{\mathrm{es}}$ = 1.42~GHz.
The energy level structure including both $|g\rangle$ and $|e\rangle$ is shown in Fig.~\ref{fig2}(c).

With a green laser excitation at 515~nm or 532~nm, an electron is non-resonantly excited above $|e\rangle$,
but is not photoionized (the band gap of diamond is 5.47~eV or 227~nm, and the NV center is a `deep' defect).
The electron relaxes back to $|g\rangle$ through spin-preserving, phonon-mediated optical transitions.
The photoluminescence spectrum of the NV center reveals a zero-phonon line at 637~nm and a broad phonon sideband in longer wavelengths extending up to 800~nm.
However, this is not the only path for the electron to return to $|g\rangle$.
There is a non-radiative (except an infrared transition at 1042~nm) channel via spin singlet states called the intersystem crossing [Fig.~\ref{fig2}(c)].
The intersystem crossing is spin-selective, and involves a spin-flip.
The electrons in the $m_{\mathrm{s}}$ = $\pm$1 states preferentially decay into this channel, and end up in the $m_s$ = 0 state of $|g\rangle$.
This provides a convenient means to read out and initialize NV spins.
By irradiating a green laser (optical pumping) and monitoring how many photons are emitted from the NV center, we are able to determine the spin state of the NV center.

A typical sample we use is an electronic-grade synthetic diamond substrate available from Element Six.
It is specified to contain less than 5~ppb nitrogen impurities and less than 0.03~ppb NV centers (either neutral or negatively-charged).
The nitrogen impurities primarily occupy the substitutional sites (called P1 centers) serving as donors, but other forms of nitrogen-related defects abundantly exist.
By noting that the number density of carbon atoms in diamond is 1.77 $\times$ 10$^{23}$~cm$^{-3}$,
the 0.03~ppb of NV centers correspond to 5 $\times$ 10$^{12}$~cm$^{-3}$, or 5 NV centers/$\mu$m$^3$.
On the other hand, with a standard (often home-built) scanning confocal microscopy setup,
one can achieve the diffraction-limited resolution ($<$ 500~nm) in lateral directions and the vertical resolution of about 1~$\mu$m without much hassle.
Therefore, there is a good chance of resolving fluorescence from a single NV center,
and in fact, we typically collect photons from a single NV center at the rate of a few tens kilo-counts per second (kcps)
using a commercial single-photon counting module (SPCM, from Excelitas Technologies or Laser Components).

With this ability to resolve a single NV center, an experimental setup for ODMR of the NV centers is quite different from the one for standard electron paramagnetic resonance (EPR) at X-band ($\sim$10~GHz)
utilizing a resonant cavity to uniformly and strongly drive a spin ensemble.
In the latter, a small change in the cavity quality factor upon microwave absorption by the spin ensemble is detected.
In ODMR, the emitted optical photons carry the information of the NV spin state.
The microwave absorption that drives the NV spin state from $m_{\mathrm{S}}$ = 0 to $\pm$1 is signaled by the reduction in the photon counts (by 30~\% in the best case).
Correspondingly, requirements for the uniformity of the microwave field and the homogeneity of the static magnetic field are greatly relaxed,
since we have only to drive a single NV center which is spatially localized (the bond length of diamond is 0.154~nm).
We may use a position-controlled permanent magnet to supply $B_0$.
A most common way to generate an oscillating magnetic field $B_1$ has been to use a thin straight metal (copper or gold) wire placed across a diamond sample.
This is very simple and highly broadband.
One can even use the same metal wire to drive nuclear spins to perform electron-nuclear double resonance.
In ODMR of the NV center, the microwave frequency $f_{\mathrm{mw}}$ is swept under fixed $B_0$,
whereas in standard EPR $B_0$ provided by an electromagnet is swept with $f_{\mathrm{mw}}$ fixed at the cavity resonance.

Before proceeding, we briefly mention some of nuclear species we will be interested in in the later sections [Table~\ref{tab_nuc}].
\begin{table}
\caption{\label{tab_nuc}Nuclei of interest. $\gamma_{\mathrm{n}}$: nuclear gyromagnetic ratio, NA: natural abundance.}
\begin{ruledtabular}
\begin{tabular}{ccdd}
& $I$ & \multicolumn{1}{c}{$\gamma_{\mathrm{n}}$ (kHz/mT)} & \multicolumn{1}{c}{NA (\%)} \\
\hline
$^{12}$C & 0 & 0 & 98.93 \\
$^{13}$C & 1/2 & 10.705 & 1.07 \\
\hline
$^{14}$N & 1 & 3.077 & 99.63 \\
$^{15}$N & 1/2 & -4.316 & 0.37 \\
\hline
$^{1}$H & 1/2 & 42.577 & 99.99 \\
\end{tabular}
\end{ruledtabular}
\end{table}
Diamond is primarily composed of nuclear-spin-free $^{12}$C isotopes.
Still, a small fraction of $^{13}$C isotopes with $I$ = 1/2 cause decoherence of the NV electronic spin, and isotopic purification is often conducted to improve the spin coherence.
As will be discussed in Sec.~\ref{sec_ms}, we create near-surface NV centers by either N$^+$ ion implantation or nitrogen-doping during CVD growth.
Yet, diamond substrates on which N$^+$ ion implantation or CVD is conducted already contain NV centers.
To distinguish between native (unintentional) and artificial (intentional) NV centers, $^{15}$N isotopes, which are scarce in the former, are often used for the latter.
The different nuclear spins of $I$ = 1 for $^{14}$N and $I$ = 1/2 for $^{15}$N result in hyperfine triplet and doublet splittings with $^{14/15}$NV electronic spins, respectively.
Interestingly, the signs of the nuclear gyromagnetic ratios $\gamma_{\mathrm{n}}$ are also different, which will turn out to be useful in interpreting AC magnetometry data in Sec.~\ref{sec_nuclear}. 

\section{Microwave engineering\label{sec_me}}
The use of a metal wire for ODMR is convenient, but there are some shortcomings.
The main concern is that $B_1$ decreases rapidly as departing from the wire.
The spin resonance is induced only when the NV centers are located in the vicinity of the wire.
This is particularly inconvenient when we search for single NV centers within a wide range of a diamond substrate, which is typically a few mm$^2$ in area. 
Moreover, the wire is placed in contact with a diamond surface, nearby which an objective lens and/or a specimen to be sensed are present.
One cannot observe, for instance, the region beneath the wire, as it hinders the light propagation.
It is preferable to leave the surface open as much as possible.
For wide-field magnetic imaging using NV ensembles, the uniformity of the $B_1$ field becomes more important.

\begin{figure}
\begin{center}
\includegraphics{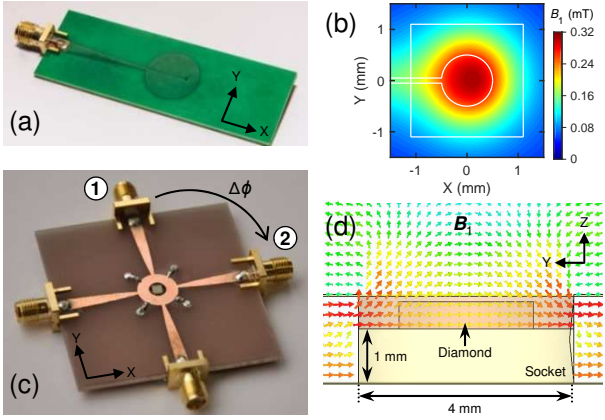}
\caption{(a) A photograph of a broadband, large-area microwave antenna.
(b) Simulated $B_1$ at the diamond surface (the white square represents the edges of the 2~$\times$~2~$\mu$m$^2$ sample)
at the microwave frequency $f_{\mathrm{mw}}$ = 2.87~GHz and the microwave power $P_{\mathrm{mw}}$ = 1~W.
(c) A photograph of polarization-tunable microwave circuit.
(d) Simulated vectorial distribution of $B_1$ in the YZ plane at $f_{\mathrm{mw}}$ = 2.87~GHz.
(a), (b) The original data presented in Ref.~\onlinecite{SMS+16}.
(c), (d) The original data presented in Ref.~\onlinecite{HAS+16}. 
\label{fig3}}
\end{center}
\end{figure}

Figure~\ref{fig3}(a) shows a microwave planar ring antenna designed to address these limitations.
It is designed to be placed beneath a diamond sample, to have a resonance frequency at around 2.87~GHz with the bandwidth of 400~MHz,
and to generate $B_1$ spatially uniform within an circular area with a diameter of 1~mm [Fig.~\ref{fig3}(b)].
The direction of $B_1$ is primarily along the $Z$ direction (perpendicular to the surface).
This antenna has been instrumental in quick search of NV centers to characterize CVD-grown diamond samples.

The second example of microwave engineering is more fundamental.
In ideal magnetic resonance experiments, the directions of $B_1$ and $B_0$ should be orthogonal.
On the other hand, the direction of $B_1$ generated by a wire is highly position-dependent and there is no guarantee that it is orthogonal to that of $B_0$ at the location of the NV center.
Moreover, $B_1$ from a wire is linearly polarized.
On the other hand, the $m_S$ = 0 $\leftrightarrow$ 1 ($-$1) transition is driven by $\sigma^{+}$ ($\sigma^{-}$) circularly polarized microwaves [Fig.~\ref{fig2}(b)].
Of course, the linear polarization is a superposition of $\sigma^{\pm}$, and the both transitions are driven non-selectively by linearly polarized microwaves.
Nonetheless, to fully exploit the $S$ = 1 nature of the NV spin, we need a reliable means to generate arbitrarily polarized microwave magnetic fields.

Figure~\ref{fig3}(c) is a planar microwave circuit providing spatially uniform, arbitrarily polarized, and {\it in-plane} microwave magnetic fields.
With a (111)-oriented diamond and $B_0$ applied perpendicular to the diamond surface,
both $\bm{B}_0 \perp \bm{B}_1$ and $\bm{n}_{\mathrm{NV}} \perp \bm{B}_1$ are realized.
The microwaves are fed through two orthogonal ports with the phase difference $\Delta \phi$, which controls the microwave polarization.
The resonance frequency of the circuit can be adjusted by the four chip-capacitors mounted on the four points of the circuit.
By using varactor diodes as variable chip capacitors, the resonance frequencies can be tuned between 2~GHz and 3.2~GHz. 

\begin{figure}
\begin{center}
\includegraphics{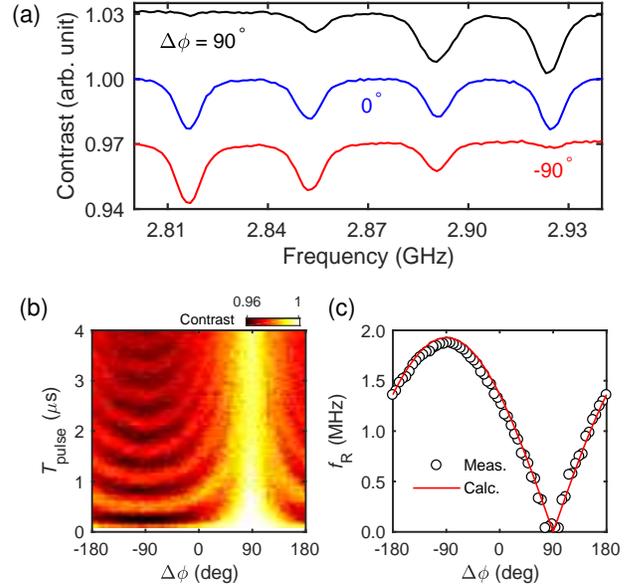}
\caption{(a) CW ODMR spectra at $\Delta \phi$ = 90$^{\circ}$ (black, shifted by +0.03), 0$^{\circ}$ (blue) and $-$90$^{\circ}$ (red, shifted by $-$0.03).
(b) Rabi oscillations as a function of $\Delta \phi$ at $P_{\mathrm{mw}}$ = 4.56~W.
(c) Rabi frequencies $f_{\mathrm{R}}$ as a function of $\Delta \phi$.
The solid line is a calculation based on three-dimensional electromagnetic simulations.
The original data presented in Ref.~\onlinecite{HAS+16}.
\label{fig4}}
\end{center}
\end{figure}

Figure~\ref{fig4}(a) shows three representative continuous-wave (CW) ODMR spectra of ensemble NV centers in type-Ib (111) diamond.
The `NV ensemble' contains NV centers oriented along the [111], [1$\bar{1}\bar{1}$], [$\bar{1}$1$\bar{1}$], and [$\bar{1}\bar{1}$1] directions with equal probabilities
(a notable exception is the NV ensemble perfectly aligned along one axis using special CVD growth conditions~\cite{EHC+12,MTZ+14,LTT+14,FDM+14}).
Not to mention, $B_0$ can be aligned with only one of the four orientations at one time.
Here, $B_0$ = 1.9~mT is aligned along [111] that is perpendicular to the surface [Fig.~\ref{fig2}(a)].
The outer two dips of the spectra are from the NV centers with $\bm{n}_{\mathrm{NV}} \parallel \bm{B}_0$,
whereas the inner two from the three degenerate, non-aligned NV centers experiencing a weaker static magnetic field of $B_0 \cos(109.5^{\circ})$.
The vertical axis is given as contrast, which represents the photon counts normalized by those under the off-resonance condition. 
At $\Delta \phi$ = 90$^{\circ}$ ($-$90$^{\circ}$), only the $m_S$ = 0 $\leftrightarrow$ 1 ($-$1) transition at $D_{\mathrm{gs}} \pm \gamma_{\mathrm{e}} B_0 \approx$ 2.923 (2.817)~GHz
is excited, corresponding to the $\sigma^{+}$ ($\sigma^{-}$) polarization.
The NV spins contributing to the inner two transitions are likely to experience elliptic polarizations.

For time-domain measurements, we set $f_{\mathrm{mw}}$ at the $m_{\mathrm{S}}$ = 0 $\leftrightarrow -$1 resonance,
and burst a microwave pulse for the duration of $T_{\mathrm{pulse}}$.
As increasing $T_{\mathrm{pulse}}$, the NV spin state oscillates between $m_{\mathrm{S}}$ = 0 and $-$1: Rabi oscillation. 
The entire experimental sequence is given as $\tau_{\mathrm{I}} - T_{\mathrm{pulse}} - \tau_{\mathrm{R}}$,
where $\tau_{\mathrm{I, R}}$ are the durations of green laser excitation for spin initialization and readout.
An SPCM is time-gated and collects optical photons for the duration during which the photon counts from $m_{\mathrm{S}}$ = 0 and $-$1 are appreciably different (typically $<$ 500~ns).

Rabi oscillations as a function of $\Delta \phi$ are color-plotted in Fig.~\ref{fig4}(b).
Rabi frequencies $f_{\mathrm{R}}$ extracted from Fig.~\ref{fig4}(b) are plotted in Fig.~\ref{fig4}(c).
It is observed that the oscillation at $\Delta \phi$ = $-$90$^{\circ}$ is $\sqrt{2}$-times faster than that at $\Delta \phi$ = 0$^{\circ}$.
This is because the linearly polarized microwave at $\Delta \phi$ = 0$^{\circ}$, being a superposition of $\sigma^{\pm}$,
can use only half of its energy to drive the $m_{\mathrm{S}}$ = 0 $\leftrightarrow$ $-$1 transition.
At $\Delta \phi$ = 90$^{\circ}$, the $\sigma^{+}$-polarized microwave cannot drive the $m_{\mathrm{S}}$ = 0 $\leftrightarrow$ $-$1 transition, and the Rabi oscillation is fully suppressed.
These results demonstrate near-perfect selective excitation of the NV spins.

\section{Materials science\label{sec_ms}}
For magnetometry, the proximity of the NV sensor to a specimen is crucial, as their dipolar coupling strength decays as the inverse cube of the separation [Fig.~\ref{fig1}(a)].
We thus want to create the NV centers close to the surface.
In this section, we first outline how high quality `bulk' diamond is synthesized (Sec.~\ref{sec_growth})
and then discuss how to introduce the NV centers near the surface of this host diamond by N$^+$ ion implantation (Sec.~\ref{sec_ion}) or nitrogen-doping during CVD (Sec.~\ref{sec_cvd}).

\subsection{Diamond growth techniques\label{sec_growth}}
There are primarily two kinds of synthetic diamond, namely HTHP (high-temperature high-pressure) diamond and CVD diamond.

The HTHP process of diamond synthesis aims to create conditions similar to those of the earth's mantle, where natural diamond is created.
At ambient conditions, diamond is a metastable allotrope of carbon, since graphite is energetically more stable.
Nonetheless, once formed in the HTHP environment, diamond is not readily transformed into graphite even after it is brought to the ambient condition, owing to the large activation energy.
Typical HTHP diamond contains high concentration of nitrogen (on the order of 100~ppm or more),
but high quality single crystal HTHP diamond containing nitrogen less than 0.1~ppm has also been obtained, for instance, by Sumitomo Electric Industries.~\cite{SS96}

In the case of CVD, diamond is synthesized under highly non-equilibrium conditions where a mixture of methane (CH$_4$) and hydrogen (H$_2$) gases exists as plasma.
The plasma environment is created using a vacuum chamber forming a microwave cavity at 2.45~GHz (system from Seki-ASTeX).
A step-flow growth leading to single crystalline quality has been realized under the conditions of, for instance,
the microwave power of 750~W, the pressure of 25~torr, and the substrate temperature of 800~$^{\circ}$C.~\cite{WTY+99}
Not only can CVD diamond be made chemically pure (such as the Element Six diamond mentioned in Sec.~\ref{sec_basic}),
but be isotopically pure by using $^{12}$CH$_4$ methane source.~\cite{IW14,TYW+15}
The gas purified up to 99.999~\% is commercially available (from Cambridge Isotope Laboratories).

\subsection{N$^+$ ion implantation\label{sec_ion}}
Ion implantation is the process in which accelerated charged elements hit and penetrate into a target material.
A large number of elements can be used, and the typical implantation energy ranges from 10~keV to a few MeV and the dose (fluence) from 10$^{11}$~cm$^{-3}$ to 10$^{16}$~cm$^{-3}$.
Inherent to this technique is that it introduces not only the elements of interest but also damages such as vacancies into the target material.
An additional annealing process to relax the damages is thus required.

\begin{figure}
\begin{center}
\includegraphics{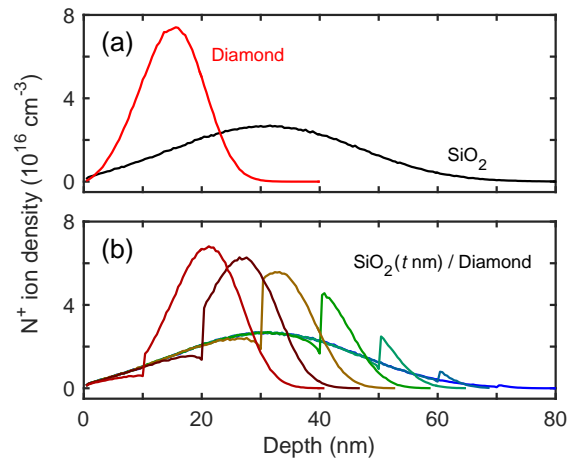}
\caption{Monte Carlo simulations of (a) N$^+$ ion implantations into diamond and SiO$_2$, and (b) SiO$_2$ ($t$~nm)/diamond with $t$ varied from 10 to 70~nm with a 10-nm step.
The N$^+$ energy is 10~keV and the dose is 10$^{11}$~cm$^{-2}$.
The original data presented in Ref.~\onlinecite{ISS+17}.
\label{fig5}}
\end{center}
\end{figure}

For the creation of NV centers, it is convenient to have nitrogen atoms and vacancies at the same time.
The annealing process promotes the diffusion of vacancies, and they subsequently pair up with nitrogen atoms to form NV centers.
It should be noted that the implantation is a stochastic process and the depth profile exhibits an approximately Gaussian distribution.
Figure~\ref{fig5}(a) shows a Monte Carlo simulation (by a software package SRIM) of N$^+$ ion implantation with the acceleration energy of 10~keV and the dose of 10$^{11}$~cm$^{-2}$.
The N$^+$ ions implanted into diamond are peaked at 15~nm from the surface and have the width of 5.4~nm.
Importantly, in reality, the ions can penetrate deeper into the crystal lattice than simulated due to the ion channeling effect, which this simulation does not take into account.
A prevailing approach to creating shallow NV centers is to keep the implantation energy low ($<$ 5~keV).~\cite{OPC+12}
The creation of single NV centers also requires low doses ($\sim$10$^8$~cm$^{-2}$).

Here, we discuss a simple method utilizing a screening mask to control both the depth profile and the dose.
While in principle a variety of materials can be adopted as screening masks, we employ SiO$_2$ owing to the ease of handling.
In addition, SiO$_2$, being an amorphous material, can suppress the ion channeling.
When the SiO$_2$ layers with thickness $t$~nm are added on top of diamond,
the depth profiles appear as combinations of the profiles of the two materials [Fig.~\ref{fig5}(b)];
down to the SiO$_2$/diamond interfaces at $t$~nm, the profiles trace that of SiO$_2$, and after entering into diamond the profiles mimic that of diamond with the near-surface parts truncated.
A similar truncated profile can be obtained by plasma-etching the implanted diamond surface, but
special cares must be taken in order to prevent the etching itself from damaging the surface.~\cite{CGO+15,OMW+15}
Notably, for $t \geq$ 40~nm, only the tail parts appear in the profiles inside of diamond, thereby the locations of the highest ion densities are {\it at the surface} with significantly reduced effective dose.
It is also clear that the distributions in the depth direction become much narrower than that without SiO$_2$.

\begin{figure}
\begin{center}
\includegraphics{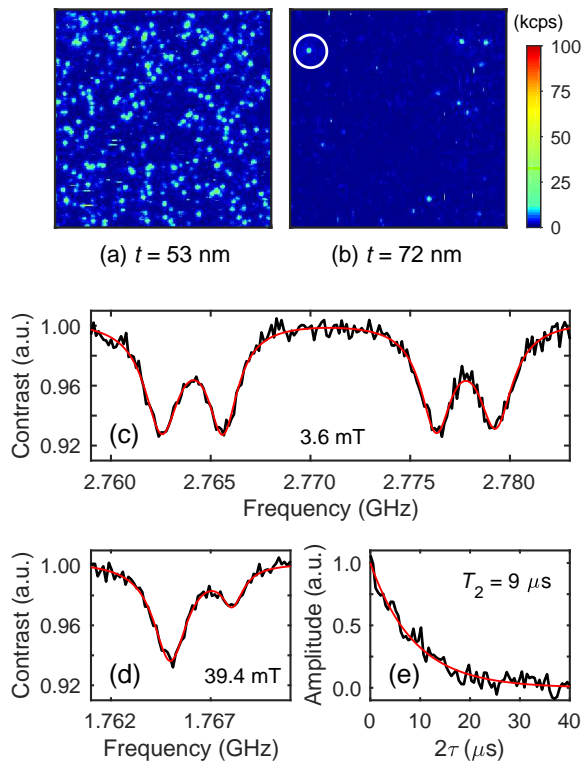}
\caption{Fluorescence images (20~$\times$~20~$\mu$m$^2$) of areas with (a) $t$ = 53~nm and (b) 72~nm.
The implantation energy was 10~keV and the does was 10$^{11}$~cm$^{-2}$.
The white circle indicates the spot from which the data in (c--e) were taken.
(c) ODMR spectrum of a single NV center, exhibiting the hyperfine couplings with both $^{13}$C and $^{15}$N nuclei.
(d) ODMR spectrum of the same NV center as in (c) at 39.4~mT.
(e) Hahn echo decay curve taken at the condition of (d).
The original data presented in Ref.~\onlinecite{ISS+17}.
\label{fig6}}
\end{center}
\end{figure}

To test this method, $^{15}$N$^+$ ion implantation is carried out onto a diamond sample on which SiO$_2$ layers with various thicknesses are deposited. 
The SiO$_2$ layers are then chemically removed, and the sample is annealed to promote the formation of NV centers.
Figures~\ref{fig6}(a,b) are exemplary fluorescence images taken at the diamond surface and from the areas with $t$ = 53~nm and 72~nm after these treatments.
The both areas show successful creations of single NV centers seen as bright spots, but with different densities reflecting the different thicknesses.

Now, we take this opportunity to explain the spin physics of the NV centers not discussed so far: coupling between the NV electronic spin and nuclear spins.
We note that, while the data we discuss below were obtained with a shallow NV center produced by the N$^+$ implantation through SiO$_2$,
the observed spectral features can be obtained from a deep NV center as well.
A particular example of ODMR spectrum of a created single NV center is shown in Fig.~\ref{fig6}(c).
At $B_0$ = 3.6~mT, we naively expect the resonance dip corresponding to the $m_{\mathrm{S}} = 0 \leftrightarrow -1$ transition to appear at $D_{\mathrm{gs}} - \gamma_{\mathrm{e}} B_0 \approx$ 2.77~GHz.
Contrary to this expectation, the observed dip splits into four.
Note that a similar ODMR spectrum is obtained at around 2.97~GHz, corresponding to the $m_{\mathrm{S}} = 0 \leftrightarrow 1$ transition (not shown).
The spectrum is interpreted as follows.
The large doublet splitting of 14~MHz is due to the hyperfine coupling with a proximal $^{13}$C nucleus, accidentally present close to this particular NV center.
Previous studies have attributed this $^{13}$C nucleus with 14~MHz hyperfine coupling to the third nearest neighbor of the vacancy.~\cite{GFK08,MNR+09}

The dips further split by 3~MHz due to the hyperfine coupling with the $^{15}$N nucleus of the NV center itself.
The hyperfine interaction between the NV electronic spin ($S$ = 1) and the $^{15}$N nuclear spin ($I$ = $\frac{1}{2}$), omitted in Eq.~(\ref{eq_spin}), is given as
\begin{equation}
A_{\parallel} S_z I_z + \frac{A_{\perp}}{2} (S^{+}I^{-} + S^{-}I^{+})
\label{eq_hf}
\end{equation}
with $A_{\parallel}$ = +3.03~MHz and $A_{\perp}$ = +3.65~MHz.~\cite{FEN+09}
The first term of Eq.~(\ref{eq_hf}) is diagonal and primarily contributes to the ODMR spectra.
It should be noted that these hyperfine constants are smaller than that of more distant third-shell $^{13}$C nuclei.
This reflects the fact that the electronic wavefunctions of $e_x$ and $e_y$ have very small probability densities at the position of the N atom.~\cite{GFK08}

Yet another spin physics appears when $B_0$ is raised close to 50~mT.
There, optically-pumped dynamic nuclear polarization (DNP) is in action.~\cite{JNB+09}
Recall that the excited state $|e\rangle$ is spin triplet with $D_{\mathrm{es}}$ = 1.42~GHz.
This suggests that at $B_0$ = $D_{\mathrm{es}}/\gamma_{\mathrm{e}}$ = 51~mT, the $m_{\mathrm{S}}$ = 0 and $-$1 sublevels of $|e\rangle$ become degenerate.
The Zeeman energies of the electronic and nuclear spins are almost equalized,
and the hyperfine interaction of the excited state plays a role of mixing the spin states, resulting in the excited state level anticrossing (ESLAC).
The hyperfine interaction of the excited state has the same form as Eq.~(\ref{eq_hf}) but is much stronger (40$-$60~MHz~\cite{DMD+13}),
because the excited state is composed of the $a_1$ state, which has a larger probability density at the position of the N atom.~\cite{GFK08}
In the $(m_{\mathrm{S}}, m_{\mathrm{I}})$ basis of the excited state, the hyperfine interaction induces the filp-flop of the spins: $(m_{\mathrm{S}}, m_{\mathrm{I}}) = (0, -\frac{1}{2}) \leftrightarrow (-1, \frac{1}{2})$.
Combined with laser illumination, the role of which is to induce the electronic spin flip from $m_{\mathrm{S}}$ = $-$1 to 0,
the system is dynamically driven as $(m_{\mathrm{S}}, m_{\mathrm{I}}) = (-1, -\frac{1}{2}) \rightarrow (0, -\frac{1}{2}) \leftrightarrow (-1, \frac{1}{2}) \rightarrow (0, \frac{1}{2})$.

Figure~\ref{fig6}(d) shows the ODMR spectrum at $B_0$ = 39.4~mT (corresponding to the dips around 2.764~GHz at 3.6~mT).
Still away from 50~mT, the imbalanced dips reveal that the DNP is under development.
Clearly, the transition at the lower frequency is being polarized and assigned as $m_{\mathrm{I}}$ = $\frac{1}{2}$.
The positive $A_{\parallel}$ of the $^{15}$N nucleus brings the $m_{\mathrm{I}}$ = $\frac{1}{2}$ state into the lower frequency.
We will see in the next section that the opposite happens in the case of $^{14}$N nuclei.

To conclude this subsection, we show the coherence time $T_2$ of the NV spin
measured by a Hahn echo sequence $\tau_{\mathrm{I}} - \pi/2 - \tau - \pi- \tau - \pi/2 - \tau_{\mathrm{R}}$,
where `$\pi$' and `$\pi/2$' denote $\pi$ and $\pi/2$ pulses, respectively, and `$\tau$' is the pulse interval.
From the decay time of the signal we determine $T_2$ to be 9~$\mu$s.
Parenthetically, the longest $T_2$ obtained from the $t$ = 72~nm area is 25.4~$\mu$s, which is relatively long for shallow NV spins.
We will revisit the meaning of $T_2$ and the Hahn echo in Sec.~\ref{sec_ac_basic}, when we discuss AC magnetometry.

\subsection{Nitrogen-doping during CVD\label{sec_cvd}}
In 99.7\%-$^{12}$C CVD diamond {\it without} intentional doping, the densities of nitrogen and other paramagnetic impurities are suppressed below 10$^{13}$~cm$^{-3}$, and 
the density of native NV centers is as low as 10$^{10}$~cm$^{-3}$.~\cite{BNT+09}
Single NV centers found there exhibit superb coherence times ($T_2$ = 1.8~ms, compared with $T_2 \sim$ 300~$\mu$s for natural abundance CVD diamond),
underscoring the importance of eliminating both environmental paramagnetic and nuclear spins.
However, these NV centers with high coherence are situated far away from the surface and distributed randomly in bulk;
it is not obvious whether the coherence can be sustained even when NV centers are brought close to the surface.

To create NV centers by intentional doping during CVD growth, N$_2$ gas is additionally introduced into the growth chamber.
The amount of N$_2$ gas is usually given by the nitrogen-to-carbon gas ratio [N/C].
For instance, with [N/C] = 0.03\%, the nitrogen impurity density of about 7~$\times$~10$^{15}$~cm$^{-3}$ is obtained,
suggesting a very small incorporation of nitrogen into a growth layer (if the gas ratio were kept in the grown layer, the nitrogen density would be 5~$\times$~10$^{18}$~cm$^{-3}$).~\cite{IFS+12}
Using this doping condition and $^{12}$C purified to 99.99\%, a 100-nm-thick nitrogen-doped surface layer was grown, 
and single NV centers (density of 5~$\times$~10$^{10}$~cm$^{-3}$) with $T_2$ = 1.7~ms have been found.~\cite{IFS+12}
While the NV density and $T_2$ are similar to those of the bulk, nominally undoped CVD diamond mentioned above, the $^{12}$C purity and the nitrogen density are noticeably different.
Moreover, the NV-to-N ratio in the grown film is also very small ($<$10$^{-5}$), indicative of an inefficient conversion of nitrogen into NV centers or a lack of vacancies to pair up with nitrogen atoms.

A further complication arises in creating shallow NV centers is that the surface of as-grown CVD diamond is terminated by hydrogen atoms,
which have negative electron affinities and expropriate electrons that are needed for NV centers to be negatively-charged.
One way is to modify the surface termination.~\cite{FSBB10}
Another approach is to heavily dope nitrogen, so that the energy level of the NV center is brought below the Fermi level due to band bending.
From a 5-nm-thick surface layer of 99.99\%-$^{12}$C diamond with the NV density of 3~$\times$~10$^{11}$~cm$^{-3}$, a single NV centers with $T_2 >$ 50~$\mu$s has been obtained.~\cite{ORW+13}
In this sample, the doping condition of [N/C] = 15\%, resulting in the nitrogen impurity density of $\sim$10$^{18}$~cm$^{-3}$, was used,
and the observed $T_2$ did not vary significantly with the surface terminations.
Therefore, $T_2$ is likely to be limited by the paramagnetic spins of nitrogen impurities.

Although these results demonstrate that nitrogen-doping during CVD is a promising route to the NV creation, we still have a long way to fully optimize the growth parameters.
Among others, we point out that the misorientation angle of the diamond substrate is known to affect the morphology of the grown surface~\cite{RYY+02,TYW+15}
and is very likely to influence the efficiencies of the N incorporation and N-to-NV conversion as well.
For instance, the (100)-oriented Element Six electronic-grade substrate is specified to have the misorientation angle $< \pm$3$^{\circ}$, which is not small in terms of growth parameters.
A systematic and comprehensive study on this aspect is still lacking.

\begin{figure}
\begin{center}
\includegraphics{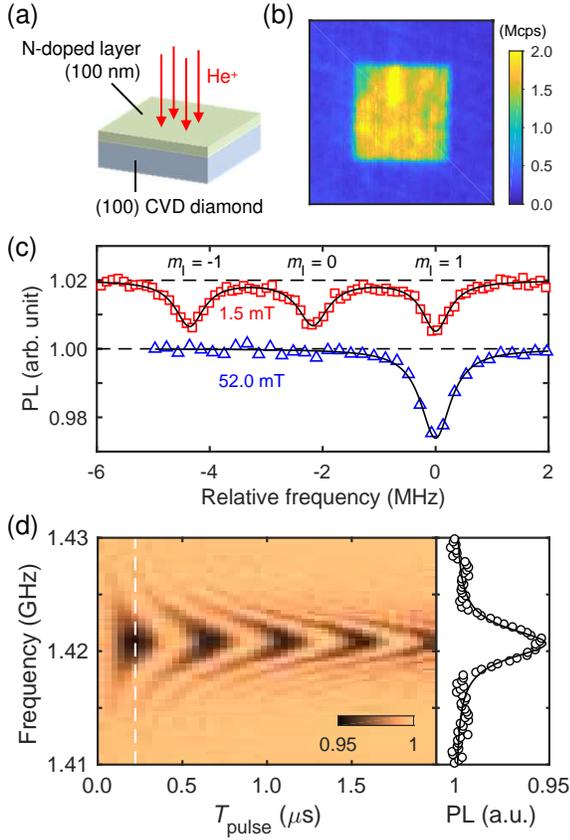}
\caption{(a) Schematic showing He$^+$ ion implantation into the diamond sample with a 100-nm-thick nitrogen-doped CVD layer on top.
(b) Fluorescence image (10~$\times$~10~$\mu$m$^2$) after He$^+$ implantation and subsequent annealing and oxidation processes.
The bright area (5~$\times$~5~$\mu$m$^2$) is the region where He$^+$ ions are implanted.
(c) CW ODMR spectra of the ensemble NV at $B_0$ = 1.5~mT (the $m_S = 0 \leftrightarrow -1$ transition, shifted by $+$0.02) and at $B_0$ = 52.0~mT.
(d) Pulsed spectroscopy of the ensemble NV at $B_0$ = 52.0~mT.
The cross section at $T_{\mathrm{pulse}}$ = 222~ns is shown on the right hand side.
(c), (d) The original data presented in Ref.~\onlinecite{SKZ+17}.
\label{fig7}}
\end{center}
\end{figure}

Even so, once nitrogen is doped, one can consider taking an additional process of creating vacancies.
There are several methods to introduce vacancies after CVD growth, for instance, through C$^+$ ion implantation~\cite{OHB+14}, electron irradiation~\cite{OHB+12}, or He$^+$ ion implantation~\cite{HLS+13}.
Here, we discuss the result of vacancy creation by He$^+$ ion implantation.
The rationale behind this approach is that light element helium causes damages less than heavier carbon,
and yet does not penetrate as deep as electrons, so that the substrate is unaffected.
We have used this technique to create a quasi-two-dimensional sheet of an NV ensemble, which is one modality of NV sensors for submicron scale magnetic imaging.
A 100-nm-thick 99.9\%-$^{12}$C surface layer with the nitrogen density of $\sim$5 $\times$ 10$^{17}$~cm$^{-3}$ is implanted by He$^+$ ions [Fig.~\ref{fig7}(a)].~\cite{KSB+16}
Figure~\ref{fig7}(b) shows a fluorescence image after He$^+$ implantation and subsequent annealing and oxidation processes.
The bright area (5~$\times$~5~$\mu$m$^2$) is the region where He$^+$ ions are implanted at the energy of 15~keV and with the dose of 10$^{12}$~cm$^{-2}$.
From the photon counts, the NV density of this area is estimated as 10$^{17}$~cm$^{-3}$, 
a significant increase compared with the NV density before He$^+$ ion implantation (on average 1.5~$\times$~10$^{15}$~cm$^{-3}$) and a highly efficient conversion from nitrogen to NV center.

Even with this high density, the linewidth is narrow, and the hyperfine splitting is clearly resolved [Fig.~\ref{fig7}(c)].
Since natural abundant nitrogen gas was used, we now observe three dips at 1.5~mT, in accordance with $I$ = 1 of $^{14}$N nuclei.
By bringing $B_0$ to 52.0~mT, $^{14}$N nuclei are fully polarized into the state with the highest frequency, contrary to the case of $^{15}$N nuclei.
This is because the hyperfine constants of the $^{14}$N nucleus are negative ($A_{\parallel}$ = $-$2.14~MHz and $A_{\perp}$ = $-$2.70~MHz).~\cite{FEN+09}
By varying $f_{\mathrm{mw}}$ around the $m_{\mathrm{I}}$ = 1 resonance, a chevron pattern typical of pulsed spectroscopy (Rabi oscillation as a function of the drive frequency) is observed [Fig.~\ref{fig7}(d)].
Such high quality near-surface NV ensembles are a promising platform for magnetic imaging.

\section{DC magnetometry\label{sec_dc}}
\subsection{Principle and sensitivity\label{sec_dc_basics}}
\begin{figure}
\begin{center}
\includegraphics{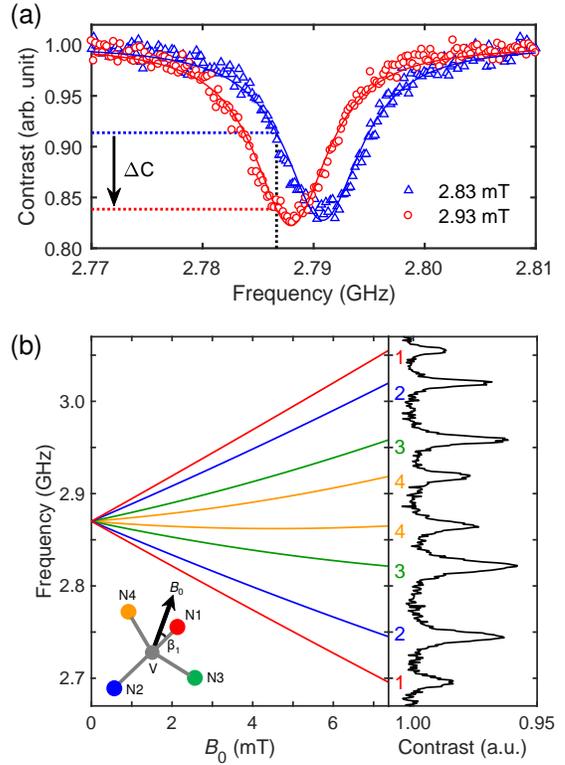}
\caption{(a) Principle of DC magnetometry.
CW ODMR spectra of a single NV center.
(b) CW ODMR spectrum of an NV ensemble for vector magnetometry (right).
Calculated $B_0$-dependence of the transition frequencies at
$\beta_1$ = 29$^{\circ}$, $\beta_2$ = 132$^{\circ}$, $\beta_3$ = 109$^{\circ}$, and $\beta_4$ = 83$^{\circ}$ (left).
(b) The original data presented in Ref.~\onlinecite{SMS+16}.
\label{fig8}}
\end{center}
\end{figure}

DC magnetometry is conceptually simple, when NV-based sensing is carried out in the CW mode.
Figure~\ref{fig8}(a) compares two CW ODMR spectra of a single NV center with the magnetic field difference of 0.1~mT.
The linewidth $\delta \nu$ is 8.2~MHz and the hyperfine splitting is not resolved.
DC magnetometry is realized as follows.
We fix $B_0$ at 2.83~mT and set $f_{\mathrm{mw}}$ at 2.7867~GHz, where the slope of the resonance dip is the steepest [the dotted vertical line in Fig.~\ref{fig8}], and keep monitoring the photon counts $I_0$.
If the magnetic field changes by $\delta B$ = +0.1~mT (which is the DC field we want to detect), we immediately notice it from the drop of the photon counts.
Noting that the slope $dI_0/dB$ is approximated by $(\gamma_{\mathrm{e}} C I_0)/\delta \nu$ with $C$ the contrast, and that the photon shot noise scales as $\sqrt{I_0}$,
we estimate the photon-shot-noise-limited DC magnetic field sensitivity $\eta_{\mathrm{sn}}$ as
\begin{equation}{
\eta_{\mathrm{sn}}^{(\mathrm{cw})} = \frac{\delta \nu}{\gamma_{\mathrm{e}} C I_0} \times \sqrt{I_0} = \frac{\delta \nu}{\gamma_{\mathrm{e}} C \sqrt{I_0}}.
\label{eq_eta_cw}}
\end{equation}
In the present case, plugging in $\delta \nu$ = 8.2~MHz, $C$ = 0.18, and $I_0$ = 30~kpcs, we obtain $\eta_{\mathrm{sn}}^{(\mathrm{cw})} \approx$ 10~$\mu$T~Hz$^{-0.5}$.
It is evident from Eq.~(\ref{eq_eta_cw}) that the DC sensitivity can be improved by achieving a narrower linewidth $\delta \nu$, a larger contrast $C$, and larger photon counts $I_0$.
Larger photon counts are attained by using NV ensembles, for which $\delta \nu$ tends to become wider due to the increased inhomogeneity.

\begin{figure}
\begin{center}
\includegraphics{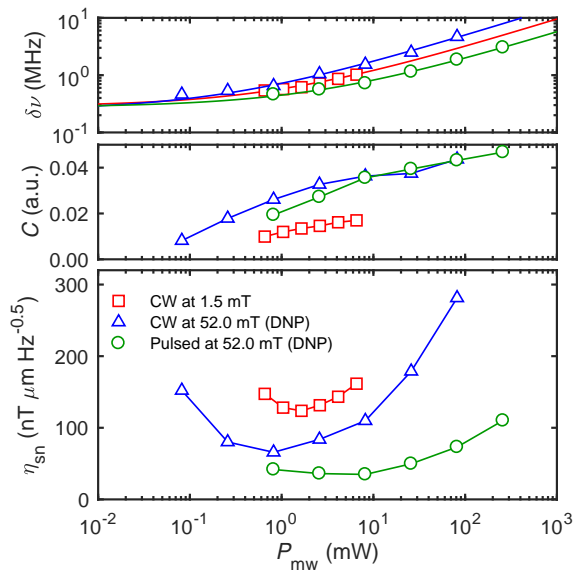}
\caption{$\delta \nu$ (top), $C$ (middle), and $\eta_{\mathrm{sn}}$ (bottom) as functions of $P_{\mathrm{mw}}$.
The solid lines in the top panel are fits to $\delta \nu_a + b P_{\mathrm{mw}}^{0.5}$.
The laser power $P_{\mathrm{L}}$ was optimized at 100~$\mu$W and 1.4~mW for CW and pulsed experiments, respectively.
The original data presented in Ref.~\onlinecite{SKZ+17}.
\label{fig9}}
\end{center}
\end{figure}

The near-surface NV ensemble discussed in Sec.~\ref{sec_cvd}, simultaneously achieving a high NV density and a narrow linewidth, is thus suited for DC magnetometry.
Here, as a concrete example, we evaluate the DC sensitivity of this NV ensemble.
In Fig.~\ref{fig9}, the squares ($\square$) are the measured $\delta \nu$ (top) and $C$ (middle) together with $\eta_{\mathrm{sn}}^{(\mathrm{cw})}$ estimated from Eq.~(\ref{eq_eta_cw}) (bottom)
as functions of the microwave power $P_{\mathrm{mw}}$.
The low-field data in Fig.~\ref{fig9} show an interplay between $\delta \nu$ and $C$.
We observe $\delta \nu = \delta \nu_a + b P_{\mathrm{mw}}^{0.5}$ with $\delta \nu_a \approx$ 250~kHz [the red solid line in the top panel of Fig.~\ref{fig9}].
The $\sqrt{P_{\mathrm{mw}}}$-dependence of $\delta \nu$ is termed as the power broadening.
The narrowing of $\delta \nu$ as decreasing $P_{\mathrm{mw}}$ is countered by the reduction of $C$,
and $\eta_{\mathrm{sn}}^{(\mathrm{cw})}$ takes its minimum at an intermediate value of $P_{\mathrm{mw}}$ = 1.64~mW.
We obtain the minimum sensitivity of 124~nT~$\mu$m~Hz$^{-0.5}$ at $B_0$ = 1.5~mT.
In ensemble-based magnetometry, it is customary to normalize the sensitivity in a unit area (1~$\mu$m$^2$) and the sensitivity unit is better given as T~$\mu$m~Hz$^{-0.5}$.

Further improvement is expected under the DNP condition, since $C$ is improved by pumping the three nuclear spin states into a single state [Fig.~\ref{fig7}(c)].
When the measurement is repeated at 52.0~mT, we obtain the minimum sensitivity of 66~nT~$\mu$m~Hz$^{-0.5}$ at $P_{\mathrm{mw}}$ = 0.82~mW [$\bigtriangleup$ in Fig.~\ref{fig9}].

In CW ODMR at a fixed $P_{\mathrm{mw}}$, increasing optical excitation power simultaneously increases $I_0$ and $\delta \nu$ while decreasing $C$.
This leads to an optimal optical power well below the saturation intensity of the NV center
(note, however, that under certain conditions with high optical power and weak microwave driving the ODMR linewidth becomes narrower~\cite{JAJB13}).
On the other hand, pulsed ODMR temporally separates the optical pumping from the spin manipulation.
A higher laser power can be used to significantly increase $I_0$ while keeping $C$ and $\delta \nu$ intact.~\cite{DLR+11}
In the example of Fig.~\ref{fig7}(d), $\delta \nu$ and $C$ can be deduced from the cross section at the $\pi$ pulse condition ($T_{\mathrm{pulse}}$ = 222~ns).
$\eta_{\mathrm{sn}}$ for pulsed ODMR is given by~\cite{DLR+11}
\begin{equation}{
\eta_{\mathrm{sn}}^{(\mathrm{pulsed})} = \frac{\delta \nu}{\gamma_{\mathrm{e}} C} \sqrt{\frac{(\pi \delta \nu)^{-1} + \tau_{\mathrm{I}} + \tau_{\mathrm{R}} }{\tau_{\mathrm{R}} I_0}}.
\label{eta_pulse}}
\end{equation}
Here, recall that $\tau_{\mathrm{I,R}}$ are the spin initialization and readout times, respectively (see Sec.~\ref{sec_me}). 
We then obtain the minimum sensitivity of 35~nT~$\mu$m~Hz$^{-0.5}$ [$\bigcirc$ in Fig.~\ref{fig9}].
These results eloquently demonstrate that magnetic resonance techniques combined with materials science have a strong impact on magnetometry.

\subsection{Vector magnetometry\label{sec_vector}}
In addition to improving DC sensitivity, NV ensembles can also be used for vector magnetometry. 
Figure~\ref{fig8}(b) shows a CW ODMR spectrum of an NV ensemble when $\bm{B}_0$ is tilted from any of the four NV axes.
In this case, the spin Hamiltonian of Eq.~(\ref{eq_spin}) is rewritten as
\begin{equation}
\mathcal{H}^{(i)} = D_{\mathrm{gs}} (S_z^{(i)})^2 + \gamma_{\mathrm{e}} \bm {B}_0 \cdot \bm{S}^{(i)},
\label{eq_spin_ens}
\end{equation}
where $\bm{S}^{(i)}$ is the $S$ = 1 spin operator with the quantization axis taken along the $i$th NV axis ($i$ = 1, 2, 3, 4).
The eight resonances are clearly separated.
The Zeeman term of Eq.~(\ref{eq_spin_ens}) is rewritten as $\gamma_{\mathrm{e}} B_0 (\sin \beta_i S_x^{(i)} + \cos \beta_i S_z^{(i)})$,
where $\beta_i$ is defined as the angle between $\bm{B}_0$ and the $i$th NV axis [the inset of Fig.~\ref{fig8}(b)].
By numerically solving Eq.~(\ref{eq_spin_ens}), we determine $B_0$ and $\beta_i$ that reproduce the eight resonances simultaneously as
$B_0$ = 7.4~mT with $\beta_1$ = 29$^{\circ}$, $\beta_2$ = 132$^{\circ}$, $\beta_3$ = 109$^{\circ}$, and $\beta_4$ = 83$^{\circ}$.
As $\beta_i$ is made closer to 90$^{\circ}$, the off-diagonal term with $S_x^{(i)}$ becomes non-negligible and the nonlinearity in the evolution of the transition frequencies is more pronounced.
When an additional small field $\bm{\delta B}$ is applied, the positions of the eight dips shift accordingly.
We can determine the vector of $\bm{B}_0 + \bm{\delta B}$ as above, and deduce the vectorial information $\bm{\delta B} = (\delta B_x, \delta B_y, \delta B_z)$.

\section{AC magnetometry\label{sec_ac}}
\subsection{Principle\label{sec_ac_basic}}
The principle of quantum sensing, on which AC magnetometry is based, is extensively discussed in Ref.~\onlinecite{DRC17},
and we only give minimum information needed to interpret the experimental data that we are going to show.
The key idea of AC magnetometry is to use quantum coherence for sensing.
As mentioned in Sec.~\ref{sec_ion}, $T_2$ is measured by a Hahn echo sequence $\tau_{\mathrm{I}} - \pi/2 - \tau - \pi- \tau - \pi/2 - \tau_{\mathrm{R}}$.
After initialization, the first $\pi/2$ pulse, say about the $x$ axis of the Bloch sphere representation of a qubit
(assigning the $m_{\mathrm{S}}$ = 0 and $-$1 (or 1) states to $|0 \rangle$ and $|1 \rangle$, respectively),
creates quantum coherence along the $y$ axis of the Bloch sphere.
The Bloch sphere corresponds to the spin vector seen from the frame of reference rotating at the resonance frequency of the NV spin,
and we assume $f_{\mathrm{mw}}$ is tuned at the resonance.
However, the NV spin experiences (quasi-)static local magnetic fields (arising from, for instance, the inhomogeneity of $B_0$ and the Overhauser field from the nuclear spin bath)
in addition to $B_0$ that defines the resonance frequency.
The spin vector then does not stay along the $y$ axis, but starts to rotate in the $xy$ plane.
Even when a single NV center is being measured, it is possible that the rotation speed and direction differ from one measurement to another.
Therefore, by averaging over many measurement runs, the signal along the $y$ axis decays with the time scale of $T_2^* \propto (\delta \nu)^{-1}$, which is often much shorter than $T_2$.

The $\pi$ pulse applied in the middle mitigates this issue.
Suppose, for a certain measurement run, the NV spin is rotating clockwise at a frequency $\delta f$ in the $xy$ plane.
After $\tau$ from the first $\pi/2$ pulse, the angle (phase) between the $y$ axis and the spin vector is $2\pi \delta f \tau$.
The $\pi$ pulse along $y$ axis brings the spin to the opposite side of the $y$ axis.
Even so, the NV spin keeps rotating clockwise, and after another duration $\tau$ it comes back to the $y$ axis, cancelling the accumulated phase.
This refocusing mechanism works irrespective of the rotating direction and the value of $\delta f$ as long as they do not change during 2$\tau$ (assumed static magnetic field).

\begin{figure*}
\begin{center}
\includegraphics{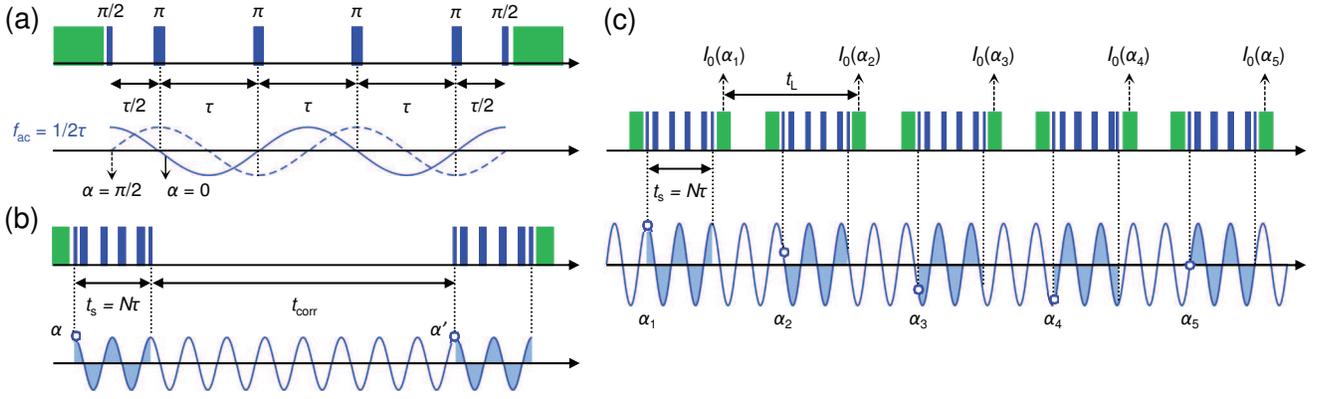}
\caption{Principles of (a) AC magnetometry, (b) correlation spectroscopy, and (c) ultrahigh resolution sensing.
\label{fig10}}
\end{center}
\end{figure*}

Now, what if $\delta f$ changes in time but periodically?
This is the situation for AC magnetometry.
The most basic protocol for AC magnetometry is shown in Fig.~\ref{fig10}(a).
The sequence is simply an extension of the Hahn echo,
but we note that the definition of $\tau$ is now the interval between the $\pi$ pulses and is not the one between the $\pi/2$ and $\pi$ pulses, which is now $\tau/2$.
The number of $\pi$ pulses, $N$, can be increased, as long as quantum coherence is preserved.
We want to detect an oscillating magnetic field of the form
\begin{equation}
B_{\mathrm{ac}}(t) = b_{\mathrm{ac}} \cos(2\pi f_{\mathrm{ac}} t + \alpha),
\label{eq_bac}
\end{equation}
where $f_{\mathrm{ac}}$ is the oscillation frequency and $\alpha$ is the phase when the initial $\pi/2$ pulse is applied.
We first consider the case of $f_{\mathrm{ac}} = (2\tau)^{-1}$ and $\alpha$ = 0 [the solid line (cosine curve) in Fig.~\ref{fig10}(a)].
Before the first $\pi$ pulse, the spin rotates clockwise and accumulates a certain phase in a nonlinear manner.
The $\pi$ pulse inverts the accumulated phase, but the rotation direction of the spin is synchronously inverted because $B_{\mathrm{ac}}(t)$ also changes its sign.
What happens is that the spin further moves away from the $y$ axis and accumulates a more phase.
This process is repeated as the $\pi$ pulses are applied.
Consequently, quantum coherence is lost quickly in this condition.
If $f_{\mathrm{ac}} \neq (2\tau)^{-1}$, multiple oscillations occurs in between the $\pi$ pulses and the phase accumulation is cancelled.
An exception is the case when $f_{\mathrm{ac}}$ satisfies $(2 k +1)/(2 \tau)$ with $k$ = 1, 2, 3...
In this case, $k + \frac{1}{2}$ oscillations occur in between the $\pi$ pulses.
$k$ oscillations cancel, but a half oscillation survives and accumulates a phase albeit smaller than the $f_{\mathrm{ac}} = (2\tau)^{-1}$ case.

Even when the condition $f_{\mathrm{ac}} = (2\tau)^{-1}$ is satisfied, the amount of the phase accumulation depends on $\alpha$.
The easiest example to appreciate this is the case of $\alpha$ = $\pi$/2 [the dashed line (sine curve) in Fig.~\ref{fig10}(a)].
Clearly, the sign of the oscillation changes in between the $\pi$ pulses, and the spin spends an equal amount of time for both signs of the oscillation;
no phase is accumulated.

The $\alpha$-dependence of the phase is calculated as~\cite{DRC17}
\begin{equation}
\varphi = \int_0^{t_{\mathrm{s}} = 4 \tau} 2 \pi \gamma_{\mathrm{e}} h(t) B_{\mathrm{ac}} (t) dt = 4\gamma_{\mathrm{e}} b_{\mathrm{ac}} t_{\mathrm{s}} \cos (\alpha).
\label{eq_phase}
\end{equation}
Here, we note that in this article the gyromagnetic ratio $\gamma_{\mathrm{e}}$ is defined in the unit of Hz/T,
and $h(t)$ describes the phase inversion by the $\pi$ pulses and is given as
\begin{eqnarray}
h(t) &=& \bigg\{
\begin{tabular}{rc}
$-$1 & for $(\frac{1}{2} - n) \tau < t < (\frac{3}{2} - n) \tau$ \\
1 & otherwise 
\end{tabular} \\
&=& \frac{4}{\pi} \sum_{n = \mathrm{odd}} \frac{\cos(\pi t/\tau)}{n}.
\label{eq_b_ac}
\end{eqnarray}
Moreover, in general, we do not have any information on $\alpha$.
The AC signal to be detected is oscillating irrespective of our measurements,
and there is no guarantee or way to phase-match the AC signal and the sensing protocol.
We are lucky in some cases and accumulate the maximum phase possible, but in other cases we are unlucky to acquire less or no phase.
We simply average over all the cases.
The probability of getting the $m_\mathrm{S}$ = 0 state at the end of the sequence in the case of `random' phases is calculated to be
\begin{eqnarray}
p_0(\tau) &=& \frac{1}{2} \left[ 1 + \frac{1}{2\pi} \int_0^{2\pi} \cos (\varphi(\alpha)) d\alpha \right] \\
&=& \frac{1}{2} [ 1 + J_0(2 \pi \gamma_{\mathrm{e}} b_{\mathrm{ac}} N \tau W_{N,\tau}) ],
\label{eq_p_0}
\end{eqnarray}
where $W_{N,\tau}(f_{\mathrm{ac}})$ is a filter function.
For a typical pulse sequence we use, $W_{N,\tau}(f_{\mathrm{ac}})$ is calculated to be
\begin{equation}
W_{N,\tau}(f_{\mathrm{ac}}) = \left| \frac{\sin(\pi f_{\mathrm{ac}} N \tau)}{\pi f_{\mathrm{ac}} N \tau} [ 1- \sec(\pi f_{\mathrm{ac}} \tau) ] \right|,
\label{eq_filter}
\end{equation}
and $J_0$ is the Bessel function of the first kind for $n$ = 0.

\begin{figure}
\begin{center}
\includegraphics{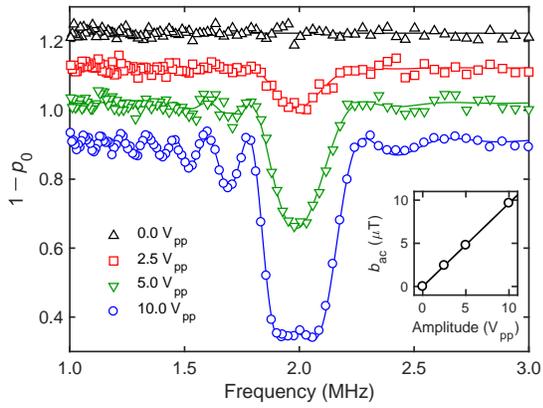}
\caption{AC magnetometry detecting an AC magnetic field at 2~MHz using a single NV center.
The data are vertically shifted for clarity.
The inset shows the relation between the input AC signal amplitude and the detected AC field strength $b_{\mathrm{ac}}$. 
\label{fig11}}
\end{center}
\end{figure}

Here, we test the sensing protocol using a single NV center in bulk diamond (Element Six electronic-grade, natural abundance).
An artificial AC magnetic field is generated by a hand-wound coil placed nearby the ODMR setup.
The coil is connected to a function generator.
When the frequency of the signal is 2~MHz, we obtain the data in Fig.~\ref{fig11}.
For a given experimental delay between $\pi$ pulses, $\tau$, we record the fluorescence signals (reflecting the loss of coherence) and convert them into $(1-p_0)$.
The sequence is repeated for different $\tau$, and $(1-p_0)$ is plotted as a function of $(2\tau)^{-1}$.
As we increase the amplitude of the signal, the dip appears at 2~MHz associated with wiggles around it.
These wiggles arise from the expression of $W_{N,\tau}(f_{\mathrm{ac}})$ given in Eq.~(\ref{eq_filter}),
which defines the `detection window' of our sensor.
We can fit the data precisely using Eq.~(\ref{eq_p_0}) [the solid lines in Fig.~\ref{fig11}] and deduce the amplitude of the signal $b_{\mathrm{ac}}$ at the position of the NV center [the inset of Fig.~\ref{fig11}].

\subsection{Nuclear spin sensing\label{sec_nuclear}}
\begin{figure}
\begin{center}
\includegraphics{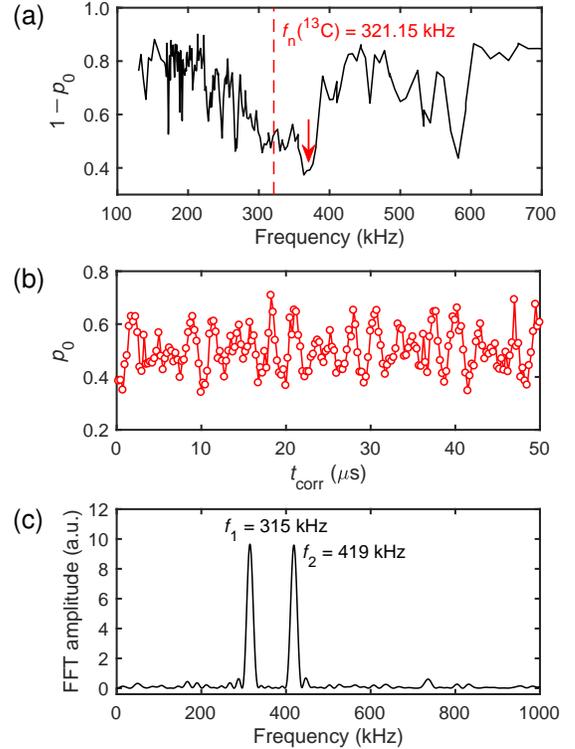}
\caption{Nuclear spin sensing by a single NV center.
(a) AC magnetometry at $B_0$ = 30.0~mT.
The dashed line indicates the Larmor frequency of $^{13}$C nuclei.
The arrow indicates the condition on which correlation spectroscopy in (b) is conducted.
(b) Correlation spectroscopy.
(c) FFT of (b).
\label{fig12}}
\end{center}
\end{figure}

Having confirmed that the protocol works, we proceed to measure nuclear spins.
We begin with standard AC magnetometry experiments, which often yield surprisingly complex spectra, and proceed to use correlation spectroscopy to understand the contents of these spectra.
We use the same single NV center used above, but of course without the application of the 2~MHz signal.
The first nuclei to look at are $^{13}$C nuclei, which are abundant in diamond.
Figure~\ref{fig12}(a) is the AC magnetometry spectrum ranging from 120~kHz to 700~kHz.
$B_0$ is set at 30.0~mT, and the corresponding Larmor (precession) frequency of $^{13}$C nuclei is 321.15~kHz [the dashed line in Fig.~\ref{fig12}(a)].
We observe a deep and broad dip structure around this frequency, but there are also several sharper dips.
In this experiment, the step of $\tau$ was set as 156~ns, and this limits the achievable resolution.
It is seen that the resolution becomes worse at higher frequencies, because the equally spaced time-domain data set is inverted to get the frequency-domain spectrum.
Yet, the whole measurement takes about a day.
Using a finer step of $\tau$ would require a dauntingly long measurement time.
Correlation spectroscopy is a complementary way to achieve a higher resolution~\cite{LDB+13,KSD+15,SRP+15,BCA+16},
and we now use it at carefully chosen frequencies in order to better understand the $^{13}$C spectrum.

The principle of correlation spectroscopy is depicted in Fig.~\ref{fig10}(b).
The two sensing sequences are separated by a time interval of $t_{\mathrm{corr}}$.
Intuitively, if $t_{\mathrm{corr}}$ is an integer multiple of $f_{\mathrm{ac}}^{-1}$,
the second sequence restarts the sensing in phase with the first sequence (it is easy to imagine the case of $\alpha$ = 0, as illustrated in Fig.~\ref{fig10}(b)).
The acquired signal is then constructive.
On the other hand, if $t_{\mathrm{corr}}$ is a half-integer multiple of $f_{\mathrm{ac}}^{-1}$, the sign of two phases are opposite.
Consequently, as $t_{\mathrm{corr}}$ is swept, the signal oscillates at the frequency of $f_{\mathrm{ac}}$.
The reason for its high resolution is that, at the end of the first sequence, the $\pi/2$ pulse brings the spin state along the $z$ axis of the Bloch sphere.
The limiting time scale of preserving the phase information is then $T_1$, not $T_2$.
$t_{\mathrm{corr}}$ can then be increased up to $T_1$ (similar to the case of stimulated echoes in magnetic resonance).

To demonstrate correlation spectroscopy, we first determine which part of the spectrum to look at.
As an example, we select the dip at 381.3~kHz [the arrow in Fig.~\ref{fig12}(a)], corresponding to $\tau$ = 1.311~$\mu$s.
Fixing $\tau$ at this value for both the first and second sequences of Fig.~\ref{fig10}(b), we sweep $t_{\mathrm{corr}}$.
The result is shown in Fig.~\ref{fig12}(b), exhibiting no decay of the oscillation amplitude up to 50~$\mu$s.
The two components, one at $f_1$ = 315~kHz and the other at $f_2$ = 419~kHz, are clearly resolved in the FFT spectrum [Fig.~\ref{fig12}(c)].
$f_1$ agrees well with $f_{\mathrm{n}}$($^{13}$C).
$f_2$ is shifted from $f_1$ by 104~kHz, interpreted as due to the hyperfine interaction between a particular $^{13}$C nucleus and the NV electronic spin.
Physically, this is understood as follows.
In the $xy$ plane of the Bloch sphere, the NV spin is in a superposition of the $m_{\mathrm{S}}$ = 0 and $-$1 (or 1) states.
The $m_{\mathrm{S}}$ = 0 component does not feel the hyperfine field from $^{13}$C, giving $f_1$,
whereas the $m_{\mathrm{S}}$ = $-$1 state feels it and gives $f_2$. 
The dip in Fig.~\ref{fig12}(a) appears at the average of the two: ($f_1 + f_2$)/2 = 367~kHz.
This mechanism has the same origin as that of electron spin echo envelope modulation (ESEEM), frequently observed
in electron-nuclear coupled systems including NV centers in diamond and phosphorus donors in silicon.~\cite{OHB+12,AIIY04,ATT+10}
However, in the case of $S$ = $\frac{1}{2}$, both $m_{\mathrm{S}}$ = $\pm \frac{1}{2}$ states exhibit hyperfine-induced oscillations.

Resolving the hyperfine constant of 104~kHz strongly indicates that the signal arises from a single $^{13}$C nucleus that resides in a particular position of the lattice site.
Other sharp dips would be associated with different $^{13}$C nuclei at different sites, and can in principle be analyzed by correlation spectroscopy at properly chosen $\tau$.
The spectrum in Fig.~\ref{fig12}(a) is thus specific to the single NV center we are using as a sensor, giving a `fingerprint' of the nuclear environment it experiences.
If we look at a different single NV center, we should still see a spectrum around this frequency range, but the detail can vary significantly.

\begin{figure}
\begin{center}
\includegraphics{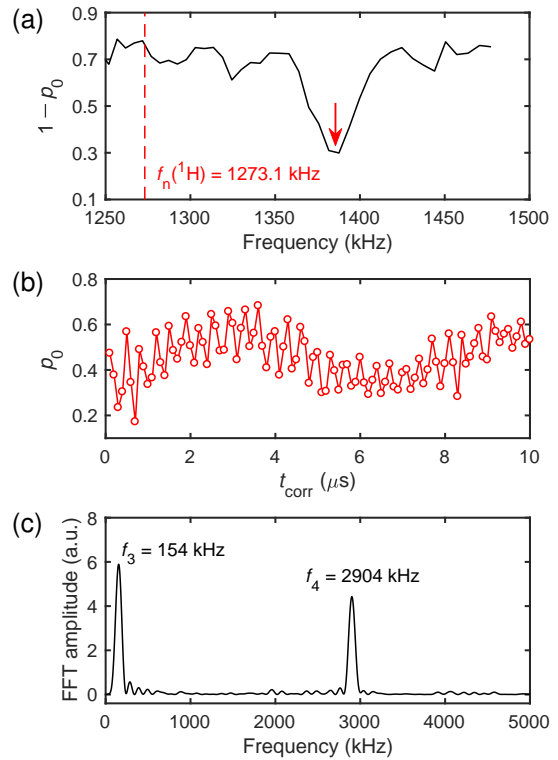}
\caption{Nuclear spin sensing by a single NV center.
(a) AC magnetometry at $B_0$ = 29.9~mT.
The dashed line indicates the Larmor frequency of $^{1}$H nuclei.
The arrow indicates the condition on which correlation spectroscopy in (b) is conducted.
(b) Correlation spectroscopy.
(c) FFT of (b).
\label{fig13}}
\end{center}
\end{figure}

In the next example, we switch the sample to the one with a shallowly-nitrogen-doped CVD-grown 99.999\%-$^{12}$C-diamond layer, hoping to detect {\it external} nuclear spins.
In this setup, an oil immersion objective lens is used for laser focusing and photon collection, and the oil contains quite a few protons.
The result of nuclear spin sensing using a near-surface single NV center is shown in Fig.~\ref{fig13}(a) with a clear dip appearing at 1370~kHz.
An apparent candidate of the origin of this dip is proton nuclear spins, of which the bear (uncoupled) Larmor frequency is $\gamma_{\mathrm{n}}$($^{1}$H) = 1273.1~kHz at $B_0$ = 29.9~mT.
If so, is the frequency difference of 100~kHz due to the hyperfine interaction between protons and the NV spin?
This hypothesis is tested by correlation spectroscopy.
Figure~\ref{fig13}(b) reveals slow and fast oscillations, the frequencies of which are identified as $f_3$ = 154~kHz and $f_4$ = 2904~kHz, respectively [Fig.~\ref{fig13}(c)].
These frequencies are quite far away from that of protons.
The correct interpretation is that these signals arise from the $^{15}$N nuclear spin of the $^{15}$NV center itself.
The bare $^{15}$N Larmor frequency is $f_{\mathrm{n}}$($^{15}$N) = $-$129.05~kHz.
Since we are detecting cosine oscillations in correlation spectroscopy, the sign of the gyromagnetic ratio cannot be discriminated.
However, if we assume $-f_3$ corresponds to the bare $^{15}$N Larmor frequency,
the difference, $f_4 - (-f_3)$ = 3.05~MHz, is exactly the hyperfine constant of the NV spin [Sec.~\ref{sec_ion}].
The dip position in Fig.~\ref{fig13}(a) is then understood as the average ($-f_3$ + $f_5$)/2 = 1374~kHz.

\begin{figure}
\begin{center}
\includegraphics{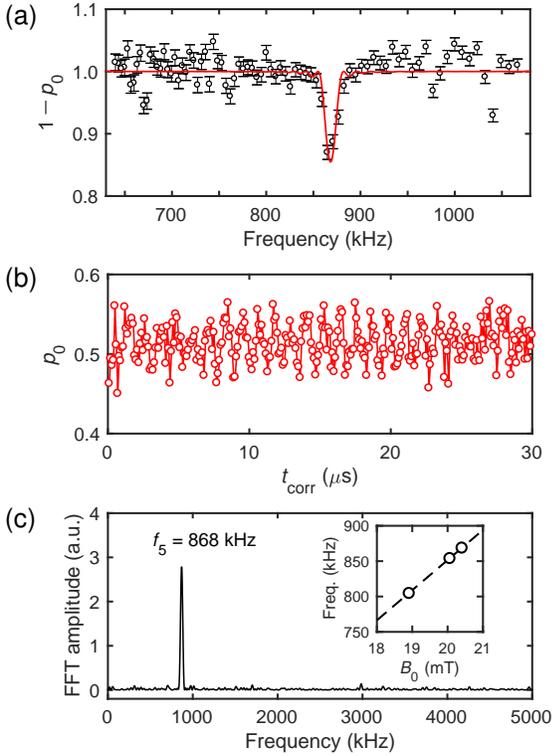}
\caption{Nuclear spin sensing by a single NV center.
(a) AC magnetometry at $B_0$ = 20.4~mT.
(b) Correlation spectroscopy.
(c) FFT of (b).
The inset is the FFT peak frequencies obtained at different $B_0$.
The dashed line is given as $f = \gamma_{\mathrm{n}}(^{1}\mathrm{H}) B_0$, and is {\it not} a fit. 
\label{fig14}}
\end{center}
\end{figure}

The final example of nuclear spin sensing, using the same diamond but a different single NV center, shows a successful detection of proton nuclear spins at the surface [Fig.~\ref{fig14}].
AC magnetometry reveals a dip exactly at the proton Larmor frequency of $f_{\mathrm{n}}$($^{1}$H) = 868~kHz at $B_0$ = 20.4~mT.
From the fit to the data based on theory described in Ref.~\onlinecite{PDC+16}, we can estimate the depth of the NV center as $d_{\mathrm{NV}}$ = 18~nm, confirming the shallow doping. 
Correlation spectroscopy also reveals the oscillation at this frequency.
As $B_0$ is changed, the change in the detected frequency exactly follows the proton Larmor frequency [the inset of Fig.~\ref{fig14}(c)].
Based on the known proton density in the oil and the detection volume of the sensor ($\propto (d_{\mathrm{NV}})^3$),
the number of protons detected here, $N_{\mathrm{H}}$, is estimated to be on the order of 10$^6$.
On the other hand, the thermal nuclear polarization at room temperature and at low magnetic fields is on the order of 10$^{-7}$,
suggesting that the number of polarized nuclear spins is less than one.
In this regime, the statistical polarization, which scales as $\sqrt{N_{\mathrm{H}}}$, exceeds the thermal polarization and produces detectable magnetic fields on the order of 100~nT .

Two important messages can be drawn from these examples of nuclear spin sensing.
The first is ``appearances are deceiving'';
signals obtained by AC magnetometry do not necessarily represent the nuclear Larmor frequencies themselves and careful analysis, such as correlation spectroscopy, is indispensable.
Recall that the odd harmonics of the detection frequency, $(2k+1)/(2\tau)$, can also be detected.
A recent study has revealed that the even harmonics $(2k)/(2\tau)$ and their odd subharmonics can also be detected.~\cite{LBR+15}
These, so-called spurious harmonics, result from the finite duration and detuning of pulses (note that $h(t)$ of Eq.~(\ref{eq_phase}) assumes an infinitesimally short pulse length and the exact resonance).
The $^{1}$H gyromagnetic ratio is coincidentally just about four times larger than that of $^{13}$C nuclei ($\gamma_{\mathrm{n}}$($^{1}$H)/$\gamma_{\mathrm{n}}$($^{13}$C) = 3.98),
and the fourth spurious harmonics of the $^{13}$C Larmor frequency can overlap with the fundamental $^{1}$H Larmor frequency.

The second issue is moderate frequency resolutions that the sensing protocols can realize.
The resolution is fundamentally limited by $T_2$ of the NV electronic spin in the case of AC magnetometry,
since there is no way to accumulate the phase after quantum coherence is lost.
The AC magnetometry spectra show even limited resolutions due to the step of $\tau$ we can use within practical measurement time.
Correlation spectroscopy enjoys better resolutions owing to the $T_1$-limited time scale.
The nitrogen nuclear spin of the NV center can be utilized as quantum memory, further improving the resolution.~\cite{APN+17,RZBD17}
There are also demonstrations of higher resolution achieved by using `quantum interpolation' and shaped pulse techniques.~\cite{ALS+17,ZSC+17}
Despite all these improvements, the achievable resolutions are still limited by physical parameters such as $T_1$ and $T_2$ of the electronic and nuclear spins.
Recently, three independent groups have demonstrated essentially the same ultrahigh resolution sensing protocol that does not rely on the spins' $T_1$ nor $T_2$.~\cite{SGS+17,BCZD17,BGL+17}

\subsection{Ultrahigh resolution sensing\label{sec_qdyne}}
The key idea of ultrahigh resolution sensing is simple and yet ingenious.
We have already discussed that in AC magnetometry the initial phase $\alpha$ is not known and we average over many runs.
The $i$th measurement picks up the information of the phase $\alpha_i$, and measured at the end of the protocol as the photon counts $I_0(\alpha_i)$.
Then, after $m$ measurements, the signal is averaged as $(1/m) \sum_i I_0(\alpha_i)$.
However, this averaging is usually carried out by repeating the same protocol many times with regular intervals $t_{\mathrm{L}}$ [Fig.~\ref{fig10}.(c)]. 
If so, the adjacent phases $\alpha_i$ and $\alpha_{i+1}$ have a definite relation given as
$\alpha_{i+1} = 2 \pi f_{\mathrm{ac}} t_{\mathrm{L}} + \alpha_{i}$,
which is the information we have discarded so far.
To utilize this information, we have only to record the time at which each measurement is done.
Even in a schematic of Fig.~\ref{fig10}(c), one can observe that the positions of $\alpha_i$ oscillate slowly,
corresponding to $f_{\mathrm{ac}} - f_{\mathrm{LO}}$ with $f_{\mathrm{LO}} = t_{\mathrm{L}}^{-1}$ (`LO' represents `local oscillator').
In reality, $f_{\mathrm{LO}}$ does not have to be set as $t_{\mathrm{L}}^{-1}$, as long as the measurement times are precisely known.
A series of time-tagged signals obtained in this way is expressed as
\begin{equation}
I \approx 4 \pi \gamma_{\mathrm{e}} b_{\mathrm{ac}} t_{\mathrm{s}} \sum_n \cos [2 \pi (f_{\mathrm{ac}} - f_{\mathrm{LO}})] n t_{\mathrm{L}} + \phi_0].
\label{eq_qdyne}
\end{equation}
FFT gives the AC signal frequency relative to $f_{\mathrm{LO}}$ (we do not have to know $\phi_0$).

The beauty of this method is that the resolution is no longer limited by physical parameters associated with the NV centers,
but limited by the accuracy and stability of the local oscillator, which are by far better than any other physical parameters we use. 
The Ulm group calls this method quantum heterodyne or `qdyne', highlighting the point that the NV center works as a mixer to combine the quantum signal
($\alpha_i$ measured by quantum coherence) and the classical LO signal.~\cite{SGS+17}
The ETH group calls it continuous sampling, since the entire measurement can be regarded as a single measurement to continuously monitor an AC oscillation.~\cite{BCZD17}
The Harvard group calls it synchronized readout, which is exactly what is done.~\cite{BGL+17}
Experimentally, the protocol itself is nothing but what we have been doing, and we have only to time-tag each measurement, which is essentially a software problem and thus straightforward to implement.
An important caveat on this protocol is that the AC signal to be detected must be coherent for the time much longer than $t_{\mathrm{L}}$ (otherwise the above-mentioned relation for $\alpha_i$ is not validated).

\begin{figure}
\begin{center}
\includegraphics{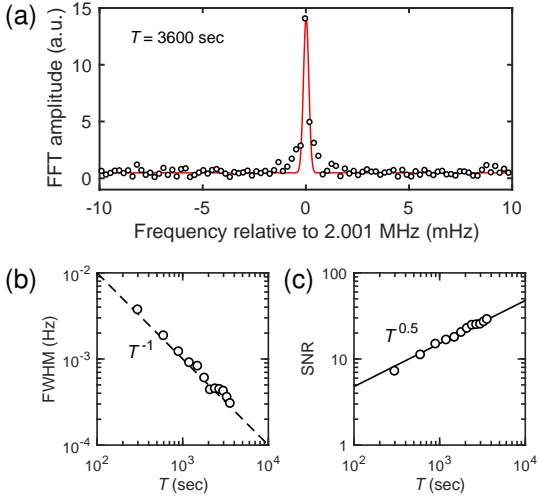}
\caption{Ultrahigh resolution sensing.
(a) FFT of the time-tagged photon count data.
The full width at half maximum (FWHM) is as narrow as 304~$\mu$Hz.
(b) FWHM (resolution) as a function of $T$.
The dashed line is given as $\mathrm{FWHM} = T^{-1}$, and is {\it not} a fit.
(c) SNR as a function of $T$.
The solid line is a fit given as $\mathrm{SNR} = 0.48 \times T^{0.5}$
\label{fig15}}
\end{center}
\end{figure}

Figure~\ref{fig15}(a) shows an experimental demonstration of this method using an artificial signal (the same setup as that of Fig.~\ref{fig11}).
After 1~hour, a phenomenal linewidth of 304~$\mu$Hz is realized.
This is a direct consequence of continuous sampling and the linewidth improves as $T^{-1}$ [Fig.~\ref{fig15}(b)].
Note that the dashed line in Fig.~\ref{fig15}(b) is {\it not} a fit, but a relation $\mathrm{FWHM }= T^{-1}$ ({\it i.e.,} a resolution of 1~Hz requires at least 1~second of measurement).
SNR scales as $\sqrt{T}$, as expected [Fig.~\ref{fig15}(c)].
A combined effect is that the precision improves as $T \times T^{0.5} = T^{-1.5}$, another notable feature of this method.
With this method, the Harvard group has been able to resolve $J$-coupling and chemical shifts of molecules placed on the diamond surface.~\cite{BGL+17}

\section{Conclusion\label{sec_conc}}
In this tutorial article, we have discussed the basics and recent development in the research of NV centers in diamond.
We have highlighted the aspects of microwave engineering, materials science, and magnetometry, and observed that they are inextricably intertwined with each other.
The research community now has basic tools to achieve high AC and DC magnetic field sensitivities and resolutions in the laboratory,
and is moving toward the goal of bringing these technologies into real and practical applications.

\section*{Acknowledgement}
EA acknowledges financial supports from KAKENHI (S) No.~26220602, JSPS Core-to-Core Program, Spintronics Research Network of Japan (Spin-RNJ),
and JST Development of Systems and Technologies for Advanced Measurement and Analysis (SENTAN).
KS acknowledges financial support from JSPS Research Fellowship for Young Scientists (DC1, KAKENHI No.~JP17J05890). 
\bibliography{tutorial_jap}
\end{document}